\documentclass[superscriptaddress,showkeys,showpacs,aps,prd,onecolumn]{revtex4}
\usepackage{amssymb,amsmath,epsfig}
\usepackage{amsmath,graphicx}
\usepackage{palatino}
\usepackage{changes}
\usepackage[Export]{adjustbox}
\usepackage[colorlinks=true, citecolor=blue, urlcolor = blue, linkcolor= red, bookmarks=true]{hyperref}
\usepackage[utf8]{inputenc}
\usepackage{listingsutf8}
\usepackage{color}     % for color figures
\usepackage{lscape}

\begin{document}

\title{\Large \bf $TeV$-Scale Resonant Leptogenesis $A_4$ with a scaling texture}

\author{H. B. Benaoum} 
\email{hbenaoum@sharjah.ac.ae}
\affiliation{Department of Applied Physics and Astronomy, \\
University of Sharjah, United Arab Emirates 
}

\begin{abstract}
We consider a TeV scale resonant lepogenesis within type-I seesaw mechanism. We show that a concrete model based on $A_4$ flavor symmetry with a scaling ansatz in the neutrino Dirac mass matrix and a retro-circulant heavy Majorana mass matrix can be realized. The full Yukawa coupling matrix $Y_{\nu}$ can be fully reconstructed from the low energy neutrino oscillations data and the quasi-degenerate heavy Majorara neutrino masses. We have carried out a detailed numerical analysis to constrain the Dirac neutrino mass matrix elements. In particular, it was found that its diagonal elements lie near or below the MeV region. Furthermore, we have investigated the allowed regions in the parameter space of the model consistent with both low energy neutrino oscillations data and resonant leptogenesis leading to the observed baryon asymmetry of the universe. Finally, the model has an imperative prediction on the allowed space for the effective Majorana neutrino mass $|m_{ee}|$ in order to account for the observed baryon asymmetry.
\end{abstract}

.

%%%%%%%%%%%%%%%%%%%%%%%%%%%%%%%%%%%%%%%%%%
\maketitle
%%%%%%%%%%%%%%%%%%%%%%%%%%%%%%%%%%%%%%%%%%

\section{Introduction}
 The observation of non-zero neutrino masses \cite{fukuda}-\cite{abe2} and baryon asymmetry of the Universe are hints for new physics beyond the Standard Model (SM) \cite{ade}. The present global analysis from several experimental data on neutrino oscillation parameters $\theta_{ij}, \Delta m_{ij}^2$ and $\delta$ can be found in \cite{capozzi,esteban}. A very elegant mechanism of generating neutrino masses as well as explaining the baryon asymmetry is leptogenesis associated with type-I seesaw \cite{minkowski}-\cite{valle}. On the other hand, the non-abelian favor discrete symmetries provide a possible underlying symmetry for the neutrino masses  \cite{king}. There are a series of such models based on the symmetry group $S_3, A_4, S_4, A_5$ and other groups with larger orders \cite{altarelli}-\cite{benaoum3}. An attractive way of realizing the observed pattern in neutrino mixing is the non-abelian discrete flavor symmetry group $A_4$ \cite{babu}-\cite{vien}. \\
The matter antimatter asymmetry is one of the long-standing problem in particle physics and cosmology. Leptogenesis has proven to be one of the most elegant way to generate the observed baryon asymmetry of the Universe.
Leptogenesis models caused by the decay of heavy Majorana right-handed neutrinos, are believed to be the underlying sources of the baryon asymmetry \cite{kuzmin}-\cite{benaoum1}. \\

Scaling ansatz in the neutrino sector has been considered by many authors and several phenomenological models based on this idea have appeared in the literature \cite{scaling}-\cite{benaoum2}. In this paper, we have investigated the neutrino mass matrix in the conventional type-I seesaw framework based on an $A_4$ flavor discrete symmetry. In \cite{benaoum2}, we have shown how to incorporate a scaling ansatz in the Dirac neutrino mass matrix with trivial form of the heavy right-handed Majorara neutrino mass matrix in an $A_4$ type-I seesaw scenario. In the present work,we consider again a model independent approach where the ratios of certain elements of the neutrino Dirac mass matrix are equal,
\begin{eqnarray}
\frac{m_{12}}{m_{33}} & = & \frac{m_{31}}{m_{22}} = \frac{m_{23}}{m_{11}} = \kappa_1 \nonumber \\
\frac{m_{21}}{m_{33}} & = & \frac{m_{13}}{m_{22}} = \frac{m_{32}}{m_{11}} = \kappa_2
\end{eqnarray}
where $m_{11}, m_{22}, m_{33}, \kappa_1$ and $\kappa_2$ are complex coefficients. \\
The Dirac neutrino mass matrix can be written as: 
\begin{eqnarray}
M_D & = & \left( \begin{array}{ccc}
m_{11} & \kappa_1 m_{33} & \kappa_2 m_{22} \\
\kappa_2 m_{33} & m_{22} & \kappa_1 m_{11} \\
\kappa_1 m_{22} & \kappa_2 m_{11} & m_{33} 
\end{array} \right)  ~~~~~~.
\label{scaling}
\end{eqnarray}
Now, instead of taking a diagonal or trivial form of the heavy right-handed neutrino mass matrix, we choose it to be a retro-circulant mass matrix,
\begin{eqnarray}
M_R & = & f~ \left( \begin{array}{ccc}
1 & q & r \\
q & r & 1 \\
r & 1 & q 
\end{array}
\right)
\end{eqnarray}
where the parameters $f$ and $r$ are considered as real numbers. \\
The light neutrino mass matrix is given by: 
\begin{eqnarray}
M_{\nu} & = & - M_D M_R^{-1} M_D^{T}  ~~~~~~.
\end{eqnarray}
In this model independent approach, the $3 \times 3$ neutrino symmetric mass matrix can be written as:
\begin{eqnarray}
M_{\nu} & = & - \frac{1}{(1- 3 q r +q^3 + r^3)} \left[(1- q r) M_{\nu~1} - (r - q^2) M_{\nu~2} + (q -r^2) M_{\nu~3} \right]
\end{eqnarray}
where, 
\begin{eqnarray}
M_{\nu~1} & = & \frac{1}{f}~\left( \begin{array}{ccc}
a^2 + 2 \kappa_1 \kappa_2~ b~ c & \kappa_2~ c^2 + (\kappa_1^2 + \kappa_2)~ a~ b & \kappa_1~ b^2 + (\kappa_1 + \kappa_2^2)~ a~ c \\
\kappa_2~ c^2 + (\kappa_1^2 + \kappa_2)~ a~ b & \kappa_2^2~ b^2 + 2 \kappa_1~ a~ c & \kappa_1 \kappa_2~ a^2 + (1+ \kappa_1 \kappa_2)~ b~ c \\
\kappa_1~ b^2 + (\kappa_1 + \kappa_2^2)~ a~ c & \kappa_1 \kappa_2~ a^2 + (1+ \kappa_1 \kappa_2)~ b~ c & \kappa_1^2~ c^2 + 2 \kappa_2 a~ b
\end{array} \right)  ~~~~~~~,
\end{eqnarray}
\begin{eqnarray}
M_{\nu~2} & = & \frac{1}{f}~\left( \begin{array}{ccc}
\kappa_2^2~ c^2 + 2 \kappa_1~ a~ b & \kappa_1 \kappa_2~ b^2 + (1+ \kappa_1 \kappa_2)~ a~ c & \kappa_2^2~ a^2 + (\kappa_1^2 + \kappa_2)~ b~ c \\
\kappa_1 \kappa_2~ b^2 + (1+ \kappa_1 \kappa_2)~ a~ c & \kappa_1^2~ a^2 + 2 \kappa_2~ b~ c & \kappa_1~ c^2 + (\kappa_1 + \kappa_2^2)~ a~ b \\
\kappa_2^2~ a^2 + (\kappa_1^2 + \kappa_2)~ b~ c & \kappa_1~ c^2 + (\kappa_1 + \kappa_2^2)~ a~ b & b^2 + 2 \kappa_1 \kappa_2~ a~ c
\end{array} \right)  ~~~~~~~,
\end{eqnarray}
\begin{eqnarray}
M_{\nu~3} & = & \frac{1}{f}~\left( \begin{array}{ccc}
\kappa_1^2~ b^2 + 2 \kappa_2~ a~ c & \kappa_1~ a^2 + (\kappa_1 + \kappa_2^2)~ b~ c & \kappa_1 \kappa_2~ 
c^2 + (1+ \kappa_1 \kappa_2)~ a~ b \\
\kappa_1~ a^2 + (\kappa_1 + \kappa_2^2)~ b~ c & c^2 + 2 \kappa_1 \kappa_2~ a~ b & \kappa_2~ b^2 + (\kappa_1^2 + \kappa_2)~ a~ c \\
\kappa_1 \kappa_2~ c^2 + (1+ \kappa_1 \kappa_2)~ a b & \kappa_2~ b^2 + (\kappa_1^2 + \kappa_2)~ a~ c & \kappa_2^2~ a^2 + 2 \kappa_1~ b~ c
\end{array} \right)  ~~.
\end{eqnarray}
with $m_{11} = a, m_{22} = c$ and $m_{33} = b$. \\

This paper is organized as follows: In section 2, we introduce the model with the scaling ansatz in neutrino Dirac mass matrix and the retro-circulant heavy Majorara Neutrino mass matrix. We show how to fully reconstruct all the elements of this matrix of the Dirac neutrino matrix $M_{\nu}$ from the low energy neutrino oscillation data. In section 3,  we explore the scenario of resonant leptogenesis for our $A_4$ model.
 We also present the numerical analysis results and investigate the viable parameter space to explain current
baryon asymmetry of the universe in section 4. In Section 5, we conclude.
%%%%%%%%%%%%%%%%%%%%%%%%%%%%%%%%%%%%%%%%%%
\section{The Model}
In order to realize the Dirac neutrino mass matrix with scaling ansatz and the heavy right-handed neutrino mass matrix with retro-circulant form, we use type-I seesaw framework based on an $A_4$ flavor discrete symmetry \cite{minkowski}-\cite{valle}. In addition to the three SM lepton $SU(2)_L$ doublets $L_{\alpha=e,\mu,\tau}$, three right-handed charged lepton singlet $e_R, \mu_R$ and $\tau_R$, and $SU(2)_L$ doublet Higgs scalar $H$, we extend the SM by three copies of right-handed singlet neutrinos $N_{i=1,2,3}$ and five $SU(2)_L$ scalar singlet fields $\phi_E, \phi_{\nu}, \xi, \xi'$ and $\xi''$. The field content of the model and their transformation properties of various fields under $A_4 \times \mathbb{Z}_3 \times \mathbb{Z}_2$ are given in Table  \ref{tab:Table 1}.\\

\begin{table}[h]
\begin{center}
\renewcommand{\arraystretch}{1.5}
\begin{tabular}{cccccccccccc}
\hline
\hline
 & $\bar{L}$ & $e_R$ & $\mu_R$ & $\tau_R$ & $N$ & $H$ & $\phi_E$ & $\phi_{\nu}$ & $\xi$ & $\xi'$ & $\xi''$\\
\hline
$SU(2)_L$  & $2$ & $1$ & $1$ & $1$ & $1$ & $2$ & $1$ & $1$ & $1$ & $1$  & 1\\
$A_4$     & $3$ & $1$ & $1'$ & $1''$ & $3$ & $1$ & $3$ & $3$ & $1$ & $1'$ & $1''$\\
$Z_3$     & $\omega$ & $\omega^2$ & $\omega^2$ & $\omega^2$ & $\omega$ & $1$ & $1$ & $\omega$ & $\omega$ & $\omega$ & $\omega$ \\
$Z_2$     & $1$ & $1$ & $1$ & $1$ & $-1$ & $1$ & $1$ & $-1$ & $1$ & $1$ & $1$ \\
\hline
\hline
\end{tabular}
\caption{\label{tab:Table 1} Fields assignments under $SU(2)_L$ and $A_4 \times \mathbb{Z}_3 \times \mathbb{Z}_2$ symmetry.}
%\label{table:Table 1}
\end{center}
\end{table}

Based on the $A_4 \times \mathbb{Z}_3 \times \mathbb{Z}_2$ symmetry, we construct the effective Lagrangian for the charged lepton Yukawa terms which is given by:
\begin{eqnarray}
{\cal L}_{cl} & = & \frac{y_e}{\Lambda} \left(\bar{L} \phi_E \right)_1 H e_R + 
\frac{y_{\mu}}{\Lambda} \left(\bar{L} \phi_E \right)_{1'}  H \mu_R + 
\frac{y_{\tau}}{\Lambda} \left(\bar{L} \phi_E \right)_{1''} H \tau_R  + H.c.
\end{eqnarray}
where $\Lambda$ represents the cutoff scale of the theory and $y_e, y_{\mu}$ and $y_{\tau}$ are respective coupling constants. \\

The Allowed terms of the neutrino invariant under $A_4 \times \mathbb{Z}_3 \times \mathbb{Z}_2$ symmetry are given by:
\begin{eqnarray}
{\cal L}_{\nu} & = & \frac{y_s}{\Lambda} \left(\phi_{\nu} \bar{L} \right)_{3_s} \tilde{H} N + \frac{y_a}{\Lambda} \left(\phi_{\nu} \bar{L} \right)_{3_a} \tilde{H} N + y_N \left(N N \right)_1 \xi + y'_N \left(N N \right)_{1'} \xi'' + y''_N \left(N N \right)_{1''} \xi' + H.c.
\end{eqnarray}
where $\tilde{H} = i \tau_2 H^*$ is the conjugate of $H$ and $y_s, y_a, y_N, y'_N$ and $y''_N$ are the coupling constants.  \\
Once the scalar $\phi_E$ gets vacuum expectation value (vev) as $<\phi_E>  =  \left(v_E, 0, 0 \right)$and the Higgs vev $<H>=v_H$ is inserted, the charged lepton mass matrix turns out to be diagonal and is given by:
\begin{eqnarray}
M_l & = & \frac{v_H v_E}{\Lambda} \left( \begin{array}{ccc}
y_e & 0 & 0 \\
0 & y_{\mu} & 0 \\
0 & 0 & y_{\tau} 
\end{array} \right)  ~~~~.
\end{eqnarray}
\subsection{Neutrino Masses and Mixings}
After the singlet scalars acquire vevs $<\xi>=u, <\xi'>=u'$  and $<\xi''>=u''$ and choosing a suitable vev alignment of the scalar triplet $\phi_{\nu}$ as: 
\begin{eqnarray}
<\phi_{\nu}>  & = & \left(v_{\nu}, v_{\nu}, v_{\nu} \right) ~~~.
\end{eqnarray}
it yields a retro-circulant right-handed heavy neutrino mass matrix,
\begin{eqnarray}
M_R & = & f~ \left( \begin{array}{ccc}
1 & q & r \\
q & r & 1 \\
r & 1 & q 
\end{array}
\right)
\end{eqnarray}
with $f = 2 y_N u, q = \frac{y'_N u''}{y_N u}$ and $r = \frac{y''_N u'}{y_N u}$. \\

The $3 \times 3$ scaling invariant Dirac neutrino mass matrix is given by: 
\begin{eqnarray}
M_D & = & \left( \begin{array}{ccc}
a & \kappa_1 a & \kappa_2 a \\
\kappa_2 a & a & \kappa_1 a \\
\kappa_1 a & \kappa_2 a & a 
\end{array} \right)
\end{eqnarray}
where $a=\frac{2}{3} y_s \frac{v_{\nu } v_H}{\Lambda}, \kappa_1 = -\frac{1}{2} - \frac{3 y_a}{4 y_s}$ and $\kappa_2 = -\frac{1}{2} + \frac{3 y_a}{4 y_s}$; with the condition $\kappa_1 + \kappa_2 = -1$. \\

The light neutrino mass matrix can be generated using type-I seesaw as,
\begin{eqnarray}
M_{\nu} & = & - M_D M_R^{-1} M_D^{T}  ~~~~~~.
\end{eqnarray}
The above matrix generates nonzero $\theta_{13}$ as well as it breaks the tri-bimaximal pattern which can be seen by rotating $M_{\nu}$ by $U_{TBM}$,
\begin{eqnarray}
U_{TBM} & = & \left( \begin{array}{ccc}
\sqrt{\frac{2}{3}} & \frac{1}{\sqrt{3}} & 0 \\
- \frac{1}{\sqrt{6}}& \frac{1}{\sqrt{3}} & -\frac{1}{\sqrt{2}} \\
- \frac{1}{\sqrt{6}} & \frac{1}{\sqrt{3}} &  \frac{1}{\sqrt{2}}
\end{array}
\right)
\end{eqnarray}
The resulting neutrino mass matrix has three eigenvalues, where one of them is zero and the others two are degenerate, which is in contradiction to the neutrino oscillation experiments. It also gives a non-vanishing reactor angle $\theta_{13}$ that deviates from tri-bimaximal mixing such that $\sin^2 \theta_{12}=1/3, \sin^2 \theta_{23} =1/2$ and $\sin^2 \theta_{13} \neq 0$. \\

In order to have the correct low energy phenomenology, we choose a more general vacuum alignment for the scalar triplet,
\begin{eqnarray}
<\phi_{\nu}>  =  \left(v_{\nu_1}, v_{\nu_2}, v_{\nu_3} \right)
\end{eqnarray}
 which can be considered as a small perturbation around the exact vev alignment. Such a choice corrects the neutrino mass spectrum and gives rise to the correct mass squared differences as well as mixing angles. The above vev configuration is a solution of the minimization conditions of the scalar potential provided it softly breaks the flavor discrete symmetry. Such a deviation is then associated to this soft breaking. The way how the above vacuum configuration is achieved is out of the scope of the present work.  \\

The chosen vevs allow us to have a scaling invariant Dirac neutrino mass matrix,
\begin{eqnarray}
M_D & = & \left( \begin{array}{ccc}
a & \kappa_1 b & \kappa_2 c \\
\kappa_2 b & c & \kappa_1 a \\
\kappa_1 c & \kappa_2 a & b 
\end{array} \right)
\end{eqnarray}
where $a=\frac{2}{3} y_s \frac{v_{\nu 1} v_H}{\Lambda}, c=\frac{2}{3} y_s \frac{v_{\nu 2} v_H}{\Lambda}, b=\frac{2}{3} y_s \frac{v_{\nu 3} v_H}{\Lambda}, \kappa_1 = -\frac{1}{2} - \frac{3 y_a}{4 y_s}$ and 
$\kappa_2 = -\frac{1}{2} + \frac{3 y_a}{4 y_s}$; with $\kappa_1 + \kappa_2 = -1$. \\

Type-I contribution to the light neutrino mass is:
\begin{eqnarray}
M_{\nu} & = & - M_D M_R^{-1} M_D^{T}  = U_{PMNS} M_{\nu}^{diag} U_{PMNS}^T~~~~~~.
\end{eqnarray}
where $M_{\nu}^{diag}=diag \left(m_1, m_2, m_3 \right)$ is the diagonal matrix containing the three neutrino masses $m_{1,2,3}$ and $U_{PMNS}= U P_{Maj}$ is the Pontecorvo-Maki-Nakagawa-Sakata unitary matrix given by: 
\begin{eqnarray}
U & = & \left( \begin{array}{ccc}
c_{12} c_{13} & s_{12} c_{13} & s_{13} e^{-i \delta} \\
- s_{12} c_{23} - c_{12} s_{23} s_{13} e^{i \delta} & c_{12} c_{23} - s_{12} s_{23} s_{13} e^{i \delta} & s_{23} c_{13} \\
s_{12} s_{23} - c_{12} c_{23} s_{13} e^{i \delta} & -c_{12} s_{23} - s_{12} c_{23} s_{13} e^{i \delta} & c_{23} c_{13} 
\end{array} \right) 
\end{eqnarray}
where $c_{ij} = \cos \theta_{ij}, s_{ij} = \sin \theta_{ij}, \theta_{ij}$'s are flavor mixing angles; $i,j=1,2,3$; $\delta$ is the Dirac CP-violating phase and $P_{Maj} = diag \left(1, e^{i \alpha/2}, e^{i \beta/2} \right)$ is the diagonal matrix that contains the Majorana phases $\alpha$ and $\beta$. \\

\subsection{Construction of the Dirac mass Matrix}
To construct the elements of the Dirac mass matrix, we decompose the neutrino mass matrix $M_{\nu}$ into two parts where the first part is a symmetric matrix and the second part is a mass matrix with zero diagonal elements having the form:
 \begin{eqnarray}
M_{\nu}^{(2)} & = & \left( \begin{array}{ccc}
0 & M_{\nu_{12}} -M_{\nu_{33}} & M_{\nu_{13}} -M_{\nu_{22}}  \\
M_{\nu_{12}} - M_{\nu_{33}}& 0 & M_{\nu_{23}} - M_{\nu_{11}} \\
M_{\nu_{13}} - M_{\nu_{22}}& M_{\nu_{23}} - M_{\nu_{11}} & 0
\end{array} \right)  = \left( \begin{array}{ccc}
0 & w_{12} & w_{13}  \\
w_{12} & 0 & w_{23} \\
w_{13} & w_{23} & 0 \end{array} \right).
\end{eqnarray}
The $w_{ij}$ are determined from the neutrino oscillation data (up to the CP-violating Majorana phases), 
\begin{eqnarray}
w_{12} & = & \sum_{i=1}^{3} \mu_i \left( U_{1i} U_{2i} - U_{3i}^2 \right) \nonumber \\
w_{13} & = & \sum_{i=1}^{3} \mu_i \left( U_{1i} U_{3i} - U_{2i}^2 \right) \nonumber \\ 
w_{23} & = & \sum_{i=1}^{3} \mu_i \left( U_{2i} U_{3i} - U_{1i}^2 \right) 
\end{eqnarray}
where we have defined the masses including the corresponding Majorana phases as $\mu_1 = m_1, \mu_2 = m_2 e^{i \alpha}$ and $\mu_3 = m_3 e^{i \beta}$. \\

Explicit structure of the off-diagonal elements of the mass matrix $M_{\nu}^{(2)}$ can be expressed as follows: 
\begin{eqnarray}
\left( \begin{array}{c} 
w_{13} \\
w_{12} \\
w_{23}
\end{array} \right) & = & \frac{1}{Det \left( M_R \right)} \left( \begin{array}{ccc}
\left(1 - \kappa_1 \kappa_2 \right) ~(r^2 -q) &  \left(\kappa_2^2 - \kappa_1 \right) (1- q r) & \left(\kappa_1^2 - \kappa_2 \right) ~(q^2 -r)  \\ 
\left(\kappa_1^2 - \kappa_2 \right) (1- q r) & \left(1 - \kappa_1 \kappa_2 \right) ~(q^2-r)  & \left(\kappa_2^2 - \kappa_1 \right) ~(r^2-q)  \\
\left(\kappa_2^2 - \kappa_1 \right) ~(q^2 -r) & \left( \kappa_1^2 - \kappa_2 \right) ~(r^2-q) & \left(1 - \kappa_1 \kappa_2 \right) (1- q r) \end{array} \right) 
\left( \begin{array}{c} 
\gamma_3 \\
\gamma_2 \\
\gamma_1 \end{array} \right)
\end{eqnarray}
where we have defined $\gamma_{i=1,2,3}$ as,
\begin{eqnarray}
\gamma_1 & = & a^2 - b c \nonumber \\
\gamma_2 & = & c^2 - a b \nonumber \\
\gamma_3 & = & b^2 - a c 
\end{eqnarray} 
which is a convenient way to parametrize the matrix elements of the Dirac mass matrix. It turns out that the parameters $\gamma_{i=1,2,3}$ can be reconstructed from the six neutrino oscillations parameters, the lightest neutrino mass, two Majorana phases, nonzero entry in the right-handed neutrino mass and scaling factors as:
\begin{eqnarray}
\left( \begin{array}{c} 
\gamma_1 \\
\gamma_2 \\
\gamma_3
\end{array} \right) & = & \frac{1}{1 - \kappa_1 \kappa_2} M_R \left( \begin{array}{c} 
w_{23} \\
w_{13} \\
w_{12}
\end{array} \right)
\end{eqnarray}

It is interesting to note that the full Dirac mass matrix, which is described altogether by 10 real parameters, can be reconstructed analytically, in the most general case, as: 
\begin{eqnarray}
a & = & \pm \frac{\gamma_1^2 - \gamma_2 \gamma_3}{\sqrt{(\gamma_1 +\gamma_2 +\gamma_3) (\gamma_1 + \omega \gamma_2 + \omega^2 \gamma_3) (\gamma_1 +\omega^2 \gamma_2 +\omega \gamma_3)}} \nonumber \\
c & = & \pm \frac{\gamma_2^2 - \gamma_1 \gamma_3}{\sqrt{(\gamma_1 +\gamma_2 +\gamma_3) (\gamma_1 + \omega \gamma_2 + \omega^2 \gamma_3) (\gamma_1 +\omega^2 \gamma_2 +\omega \gamma_3)}} \nonumber \\
b & = & \pm \frac{\gamma_3^2 - \gamma_1 \gamma_2}{\sqrt{(\gamma_1 +\gamma_2 +\gamma_3) (\gamma_1 + \omega \gamma_2 + \omega^2 \gamma_3) (\gamma_1 +\omega^2 \gamma_2 +\omega \gamma_3)}}  ~~~~~.
\label{elements}
\end{eqnarray}

The determinant of the light active neutrino mass matrix $M_{\nu}$ can now be written as: 
\begin{eqnarray}
Det \left(M_{\nu} \right) & = & \kappa_1^2 \kappa_2^2 \frac{(\gamma_1 +\gamma_2 +\gamma_3) (\gamma_1 + \omega \gamma_2 + \omega^2 \gamma_3) (\gamma_1 +\omega^2 \gamma_2 +\omega \gamma_3)}{f^3 (1+q+r)(1+ \omega q + \omega^2 r) (1+ \omega^2 q + \omega r)}
= \mu_1 \mu_2 \mu_3 
\end{eqnarray} 
The solution of the above expression determines the scaling factor $\kappa_1$ as a function of $m_{\small{lightest}}, \Delta m_{\odot}^2 , \Delta m_{atm}^2 $,
$\sin^2 \theta_{12}, \sin^2 \theta_{13}, \sin^2 \theta_{23}, \delta, \alpha$, $\beta$ and $M_R$. Detailed analysis will be performed latter on. \\
%%%%%%%%%%%%%%%%%%%%%%%%%%%%%%%%%%%%%%%%%%%%%%%%%%%%%%%%%%%

\section{Leptogenesis}
In this section, we study baryogenesis via leptogenesis of the model at TeV scale.The lepton asymmetry is generated by the CP-violating out-of-equilibrium decays of the heavy Majorana right-handed neutrinos. The lepton asymmetry is then converted by sphaleron processes into a baryon asymmetry. To realize a resonant  leptogenesis~\cite{pilaftsis}-\cite{teresi} at TeV scales in the Universe, a mass degeneracy among the heavy right-handed neutrinos is required. \\

As we will see later, the heavy Majorana mass matrix $M_R$ has a leading order parameter $f$ and two very small parameters $r$ and $q$. This is sufficient to create a very tiny splitting among the heavy Majorana masses. Although the parameters $r$ and $q$ are crucial for the resonant leptogenesis, the contribution of $r$ and $q$ in fitting the neutrino oscillation data is negligible. \\

The eigenvalues of the heavy right-handed mass matrix are given by:
\begin{eqnarray}
M_R^{diag} & = & diag \left(M_1 = f \sqrt{1-q-r-q r+q^2 +r^2}, M_2 = f (1+q+r), 
M_3 = - f \sqrt{1-q-r- q r+ q^2 + r^2} \right)  ~~~.
\end{eqnarray}
We first perform a rotation by $U_{TBM}$ on $M_R$ as follows,
\begin{eqnarray}
U_{TBM}^T M_R U_{TBM} & = & f \left( \begin{array}{ccc}
\frac{1}{2} \left(2-q-r \right) & 0 &  \frac{\sqrt{3}}{2} \left(r -q \right) \\
0 & 1+q+r & 0 \\
\frac{\sqrt{3}}{2} \left(r -q \right) & 0 & -\frac{1}{2} \left(2-q-r \right)
\end{array}
\right)
\end{eqnarray}
We note that a further rotation by a unitary matrix $U_{R~13}$ in the $13$ plane is required to diagonalize the heavy right-handed Majorana neutrino mass matrix. The full unitary matrix $U_R$ that diagonalize $M_R$ can be expressed as:
\begin{eqnarray}
U_R & = & U_{TBM} U_{R~13} 
\end{eqnarray}
where
\begin{eqnarray}
U_{R~13} & = & \left( \begin{array}{ccc}
\cos \theta_R & 0 & \sin \theta_R \\
0 & 1 & 0 \\
- \sin \theta_R & 0 & \cos \theta_R 
\end{array}
\right)
\end{eqnarray}
with 
\begin{eqnarray}
\tan 2 \theta_R = \frac{\sqrt{3} (r-q)}{2-q -r}
\end{eqnarray}
When $r=0$ and $q=0$ or $q=r$, the unitary matrix $U_R$ is the pure tri-bimaximal matrix and non-zero of the parameters $r$ and $q$ or $q=r$ encode deviation from tri-bimaximal pattern. \\

Consequently, in the basis where the heavy right-handed Majorana mass matrix is diagonal with real and positive eigenvalues, the Dirac Yukawa mass matrix $Y_{\nu}$ is found to be: 
\begin{eqnarray}
Y_{\nu} & = & \frac{1}{v_H} M_D U_R ~diag \left(1,1,i \right) \nonumber \\
& = & \frac{1}{v_H}~ \left( \begin{array}{ccc}
a & \kappa_1 d & \kappa_2 c \\
\kappa_2 d & c & \kappa_1 a \\
\kappa_1 c & \kappa_2 a & d 
\end{array} \right) 
\left( \begin{array}{ccc}
\sqrt{\frac{2}{3}} \cos \theta_R & \frac{1}{\sqrt{3}} & i~\sqrt{\frac{2}{3}} \sin \theta_R \\
- \frac{1}{\sqrt{6}} \cos \theta_R + \frac{1}{\sqrt{2}} \sin \theta_R & \frac{1}{\sqrt{3}} & -i~(\frac{1}{\sqrt{2}} \cos \theta_R + \frac{1}{\sqrt{6}} \sin \theta_R) \\
- \frac{1}{\sqrt{6}} \cos \theta_R - \frac{1}{\sqrt{2}} \sin \theta_R & \frac{1}{\sqrt{3}} & i~(\frac{1}{\sqrt{2}} \cos \theta_R - \frac{1}{\sqrt{6}} \sin \theta_R)
\end{array} \right)  ~~.
\end{eqnarray}

The CP-asymmetry parameter associated with the decay of {\em i-th} heavy Majorana neutrino into a lepton flavor $\alpha$ is defined as: 
\begin{eqnarray}
\epsilon_{N_i} & = & \sum_{\alpha}{ \frac{\Gamma \left(N_i \rightarrow \bar{L}_{\alpha} + H \right)-\Gamma \left(N_i \rightarrow L_{\alpha} + H^* \right)}
{\Gamma \left(N_i \rightarrow \bar{L}_{\alpha} + H \right)+\Gamma \left(N_i \rightarrow L_{\alpha} + H^* \right)}}
\end{eqnarray}
The general formula of the flavored CP asymmetry $\epsilon_{i}^{\alpha}$ is given by ~\cite{bhupal}:
\begin{eqnarray}
\epsilon_i^{\alpha} & = & \sum_{j \neq i} \frac{Im \left[ (Y_{\nu})^*_{\alpha i} (Y_{\nu})_{\alpha j} 
\left( Y_{\nu}^{\dagger} Y_{\nu} \right)_{ij} \right] + |\frac{M_i}{M_j}|~ 
Im \left[ (Y_{\nu})^*_{\alpha i} (Y_{\nu})_{\alpha j} \left( Y_{\nu}^{\dagger} Y_{\nu} \right)_{ji} \right]}
{\left( Y_{\nu}^{\dagger} Y_{\nu} \right)_{ii} \left( Y_{\nu}^{\dagger} Y_{\nu} \right)_{jj}} \left(f^{mix}_{ij} + f^{osc}_{ij} \right) 
\end{eqnarray}
where the regulators are: 
\begin{eqnarray}
f^{mix}_{ij} & = & \frac{\left(M_i^2 - M_j^2 \right) |M_i| \Gamma_j}{\left(M_i^2 - M_j^2 \right)^2+  M_i^2 \Gamma_j^2} \nonumber \\
f^{osc}_{ij} & = & \frac{\left(M_i^2 - M_j^2 \right) |M_i| \Gamma_j}{\left(M_i^2 - M_j^2 \right)^2 + \left(|M_i| \Gamma_i + |M_j| \Gamma_j \right)^2 
\frac{det \left[ Re \left(Y_{\nu}^{\dagger} Y_{\nu} \right)\right]}{\left( Y_{\nu}^{\dagger} Y_{\nu} \right)_{ii} \left( Y_{\nu}^{\dagger} Y_{\nu} \right)_{jj}}} ~~~~.
\end{eqnarray}
Here $\Gamma_i$ is the decay width of the {\em i-th} right-handed Majorana neutrino, defined at tree level as:  
\begin{eqnarray}
\Gamma_i & = & \frac{|M_i|}{8 \pi}\left( Y_{\nu}^{\dagger} Y_{\nu} \right)_{ii} ~~~~~~~~.
\end{eqnarray}

For each $N_i$, one may define the flavored decay parameter $K_i^{\alpha}$ as: 
\begin{eqnarray}
K_i^{\alpha} & = & \frac{\Gamma \left(N_i \rightarrow \bar{L}_{\alpha} + H \right)+\Gamma \left(N_i \rightarrow L_{\alpha} + H^* \right)}{H_N (T= |M_i|)} = 
\frac{\tilde{m}_i^{\alpha}}{m_*} 
\end{eqnarray}
where $H_N (T=|M_i|)$ is the Hubble parameter at temperature $T=|M_i|$ given by: 
\begin{eqnarray}
H_N (T=|M_i|) & = & \sqrt{\frac{4 \pi^3 g_*}{45}} \frac{M_i^2}{M_{Planck}}
\end{eqnarray}
with $g_* = 106.75$ is the effective number of relativistic degrees of freedom of the SM at high temperatures and $M_{Planck} = 1.2 \times 10^{19}~GeV$ is the Planck mass.\\
Here, the so-called equilibrium neutrino mass given by: 
\begin{eqnarray}
m_* & = & \frac{16 \pi^{5/2} \sqrt{g_*}}{3 \sqrt{5}} \frac{v_H^2}{M_{Planck}} \sim 1.08 \times 10^{-3} ~eV  ~~,
\end{eqnarray}
and the effective flavored neutrino mass $\tilde{m}_i^{\alpha}$ is defined as:
\begin{eqnarray}
\tilde{m}_i^{\alpha} & = & v_H^2 \frac{|(Y_{\nu})_{i \alpha}|^2}{|M_i|}   ~~~~~.
\end{eqnarray}
The washout factor $K_i$, which is an important ingredients for the thermodynamics description of the decays of heavy particles in the early stages of the Universe, is obtained by summing over all flavors $\alpha$, 
\begin{eqnarray}
K_i & = & \frac{\Gamma_i}{H_N (T=|M_i|)} = \sum_{\alpha} K_i^{\alpha} = \frac{\tilde{m}_i}{m_*} 
\end{eqnarray}
where $\tilde{m}_i = v_H^2 \frac{(Y_{\nu}^{\dagger} Y_{\nu})_{ii}}{|M_i|}$ is the effective neutrino mass. 
The value of the effective neutrino mass $\tilde{m}_i$ measures the departure from equilibrium and if $\tilde{m}_i << m_* (\tilde{m}_i >> m_*)$, the asymmetry is weakly (strongly) washed out by the inverse decays. 

After solving the relevant flavored Boltzmann equations and taking into account the appropriate efficiency washout and dilutions factors, the resulting baryon asymmetry of the Universe, in the strong washout regime, is estimated to be ~\cite{deppisch}: 
\begin{eqnarray}
\eta_B & = & \frac{n_B - n_{\bar{B}}}{n_{\gamma}} = - \frac{28}{51} \frac{1}{27} \frac{3}{2} \sum_{\alpha,i} \frac{\epsilon_i^{\alpha}}{K^{\alpha}_{eff} min \left(z_c, z_{\alpha} \right)}
\end{eqnarray}
where $K^{\alpha}_{eff} = \kappa^{\alpha} \sum_i K_i B_i^{\alpha}$ is the efficiency washout factor with the branching ratios $B_i^{\alpha}$ of $N_i$ decay to leptons of the $\alpha$-th flavor, 
\begin{eqnarray}
B_i^{\alpha} = \frac{|\left(Y_{\nu} \right)_{i \alpha}|^2}{\left( Y_{\nu}^{\dagger} Y_{\nu} \right)_{ii}} ~~~~,
\end{eqnarray} 
$z_c = M_1/T_c$, $T_c \sim 149~GeV$ is the critical temperature below which the sphaleron transitions freeze-out and the parameter $z_{\alpha} = 1.25 \ln (25 K^{\alpha}_{eff})$.\\

The factor $\kappa^{\alpha}$, which includes the effect of the real intermediate state subtracted collision terms, is given by: 
\begin{eqnarray}
\kappa^{\alpha} & = & 2 \sum_{i,j (j \neq i)} \frac{Re \left[ (Y_{\nu})^*_{\alpha i} (Y_{\nu})_{\alpha j} \left( Y_{\nu}^{\dagger} Y_{\nu} \right)_{ij} \right] + 
Im \left[ \left( (Y_{\nu})^*_{\alpha i} (Y_{\nu})_{\alpha j} \right)^2  \right]}
{Re \left[ \left( Y_{\nu} Y_{\nu}^{\dagger} \right)_{\alpha \alpha} \left\{ \left( Y_{\nu}^{\dagger} Y_{\nu} \right)_{ii} + \left( Y_{\nu}^{\dagger} Y_{\nu} \right)_{jj }\right\} \right]} \left(1 - 2 i \frac{|M_i| - |M_j|}{\Gamma_i + \Gamma_j} \right)^{-1} ~~~~~.
\end{eqnarray}
 %%%%%%%%%%%%%%%%%%%%%%%%%%%%%%%%%%%%%%%%%%%%%%%%%%%%%%%%%%%%%%%%%%
\section{Numerical Analysis}
In the next section, we present a detailed analysis of our work by dividing it into several subsections, first fit of the model's parameters from low energy neutrino data, compute of the baryon number asymmetry and then find the allowed space for the model. 
\label{section5}
\subsection{Parameters of  $A_4$ Model}
To perform numerical analysis in a systematical way, we use the latest global $3 \sigma$ bound on neutrino mixing angles, mass-squared differences and Dirac CP-phase. The values are shown in Table  \ref{tab:Table 2} \cite{esteban}. Notice that there are two possibilities of mass orderings allowed by neutrino experiments, the normal hierarchy (NH) $m_3 > m_2 > m_1$ and the inverted hierarchy (IH) $m_2 > m_1 > m_3$.
\begin{table}[!h]  
\begin{center}
\begin{tabular}{|c|c|c|} 
\hline
\hline  
Parameter & best fit $\pm 1 \sigma$   & $3 \sigma$ \\ 
\hline
\hline
                   & NH~~~~~IH & NH~~~~~~~IH \\
\hline
$\Delta m_{\odot}^2 [ 10^{-5} eV^2 ]$ & $7.39^{+0.21}_{-0.20}~~~7.39^{+0.21}_{-0.20}$ &  $6.79-8.01~~~6.79-8.01$ \\ 
\hline
$|\Delta m_{atm}^2| [ 10^{-3} eV^2 ]$ & $2.523^{+0.032}_{-0.030}~~~2.509^{+0.032}_{-0.030}$ & $2.432-2.618~~~2.416-2.616$ \\       
\hline
$\sin^2 \theta_{12}$               & $0.310^{+0.013}_{-0.012}~~~0.310^{+0.013}_{-0.012}$ &  $0.275-0.350~~~0.275-0.350$ \\ 
\hline
$\sin^2 \theta_{23}$               & $0.558^{+0.020}_{-0.033}~~~0.563^{+0.019}_{-0.026}$    &    $0.427-0.609~~~0.430-0.612$    \\
\hline 
$\sin^2 \theta_{13}$               & $0.02241^{+0.00066}_{-0.00065}~~~0.02261^{+0.00067}_{-0.00064}$ &  $0.02046-0.02440~~~0.02066-0.02461$ \\              
\hline
$\delta^0$                           & $222^{+38}_{-28}~~~285^{+24}_{-26}$     &   $141-370~~~205-354$             \\                                  
\hline 
\end{tabular}
\caption{\label{tab:Table 2} Global oscillation analysis with best fit for $\Delta m^2_{\odot},\Delta m^2_{atm},\sin^2 \theta_{12},\sin^2 \theta_{23},
\sin^2 \theta_{13}$ and the $\delta$ upper and/or lower corresponds to normal and/or inverted neutrino mass hierarchy.}
\end{center}
\end{table}  
Our model allows only four degrees of freedom in the Dirac neutrino mass matrix, denoted by $\kappa_1, a, b$ and $c$. To constrain the model, we perform a random scan of the inputs parameters $m_{lightest} \in [ 10^{-6}, 0.1 ]~eV, \Delta m^2_{\odot},\Delta m^2_{atm},\sin^2 \theta_{12},\sin^2 \theta_{23},
\sin^2 \theta_{13}, \delta, \alpha$ and $\beta \in [0, 2 \pi ]$. We set $f = 5~TeV$ and randomly vary the parameters $r$ and $q \in [ 10^{-8}, 10^{-10} ]$. The small variation of $r$ and $q$ helps to produce small mass splitting between the heavy Majorana neutrinos. Such a choice will generate an observed baryon asymmetry compatible with the experimental neutrino oscillation. In our analysis, we have generated one million of random sets in the $3 \sigma$ limit the current oscillation data. The correlation and constraints on the model's parameters are presented in figure 1 to figure 4 for both normal and inverted hierarchies. In figure 1, we show that the strong correlation among the parameters of the model. We obtain that the magnitude of the scaling factor $\kappa_1$ is inversely correlated to the magnitude of the parameter $a$. We also note that the magnitude of $a, b$ and $c$ are constrained in the $MeV$ range or below. It turns out that the low energy neutrino data do not favor $\kappa_1 = -1$ (i.e. $\kappa_2 =0$) which corresponds to a vanishing lightest neutrino mass as shown by the white band in the top plots of figure 1. In figure 2, we have shown the scaling factor as a function of the parameters $r$ and $q$. Interestingly, we observe that the parameter $| \kappa_1|$ takes values over the whole allowed range of $r$ qnd $q$ and this is consistent with the fact that $r$ and $q$ are needed to generate the observed baryon asymmetry. On the other hand, we found that the phase of $\kappa_1$ is restricted into three disconnected bands $\left[- 1, -0.65 \right] \pi, \left[-0.68, -0.4 \right] \pi$ and $\left[ 0.55,1 \right] \pi$. We also note that the three disconnected of $\varphi_{\kappa_1}$ are clearly seen in the top plot figure 3. For completeness, we plot in figure 3 the magnitudes versus the phases of $a, b$ and $c$ which show that all values $[ \pi, \pi]$ are allowed. \\
In Figures 4, the parameters of the model are plotted as a function of the lightest neutrino mass $m_1$ for NH and $m_3$ for IH and absolute neutrino masses where $\sum\limits_{\alpha=e,\mu,\tau} |m_{\nu_{\alpha}}|$ is ranging in between $0.055 eV$ to $0.17 eV$ for NH and $0.125 eV$ to $0.17 eV$ for IH.
 \\ 
 %%%%%%%%%%%%%%%%%%%%%%%%% fig 1 %%%%%%%%%%%%%%%%%%%%%%%%%%
\begin{figure}[hbtp]
\centering
  \begin{tabular}{@{}cc@{}}
  \includegraphics[width=.4\textwidth]{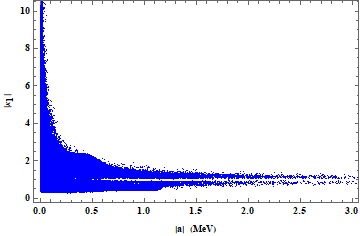} & \includegraphics[width=.4\textwidth]{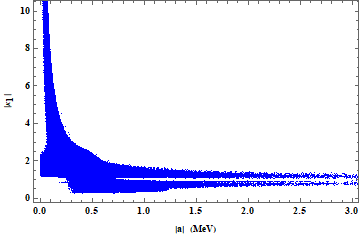} \\
    \includegraphics[width=.4\textwidth]{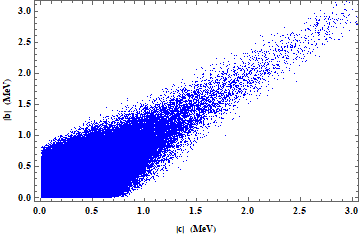} & \includegraphics[width=.4\textwidth]{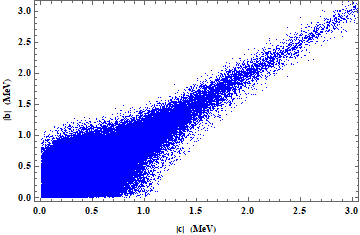} \\
\end{tabular}
\caption{ Correlation between the scaling parameter $|\kappa_1|$ and $|a|$, and correlation between the parameters $|b|$ versus $|c|$ for NH (left panel) and IH (right panel).}
\end{figure}

%%%%%%%%%%%%%%%%%%%%%%%%%%% fig 2 %%%%%%%%%%%%%%%%%%%%%%%%
\begin{figure}[hbtp]
\centering
  \begin{tabular}{@{}cc@{}}
  \includegraphics[width=.4\textwidth]{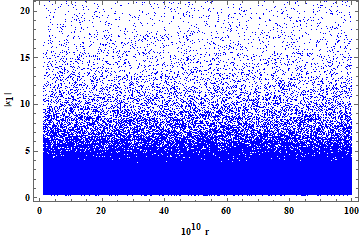} & \includegraphics[width=.4\textwidth]{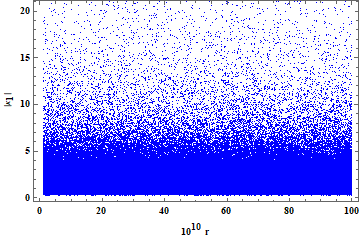} \\
 \includegraphics[width=.4\textwidth]{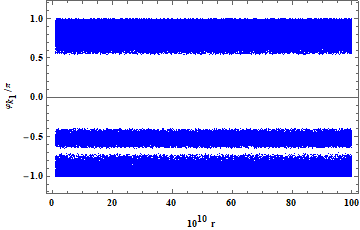} & \includegraphics[width=.4\textwidth]{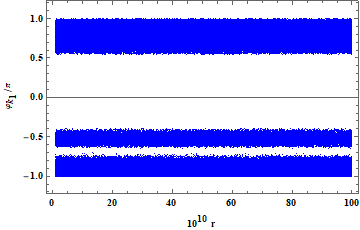}  \\
\includegraphics[width=.4\textwidth]{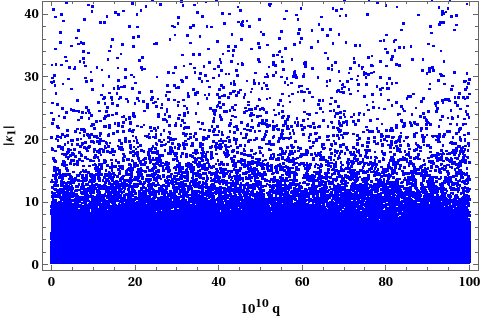} & \includegraphics[width=.4\textwidth]{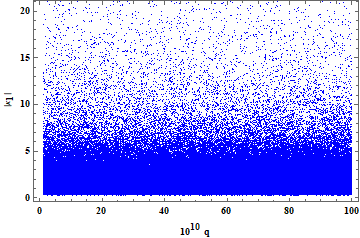} \\
 \includegraphics[width=.4\textwidth]{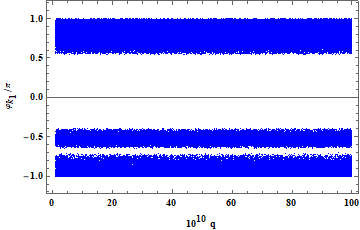} & \includegraphics[width=.4\textwidth]{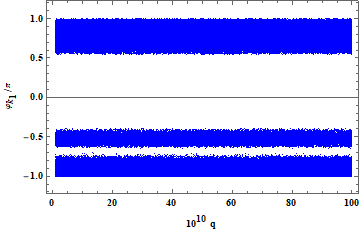}  \\
\end{tabular}
\caption{Plot of the scale factor $\kappa_1$ versus the parameters $r$ and $q$ for NH (left panel) and IH (right panel).}
\end{figure}

%%%%%%%%%%%%%%%%%%%%%%%% fig 3 %%%%%%%%%%%%%%%%%%%%%%%%%%%
\begin{figure}[hbtp]
\centering
  \begin{tabular}{@{}cc@{}}
  \includegraphics[width=.4\textwidth]{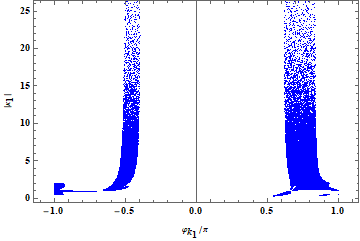} & \includegraphics[width=.4\textwidth]{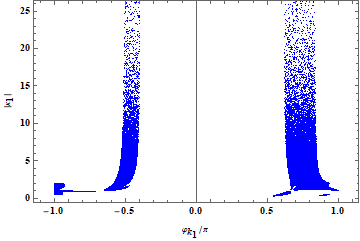} \\
  \includegraphics[width=.4\textwidth]{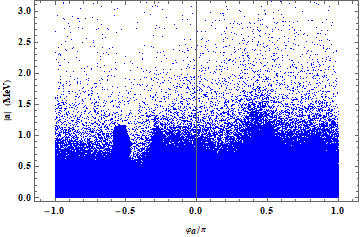} &  \includegraphics[width=.4\textwidth]{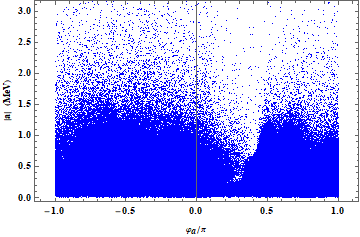}\\                                          
    \includegraphics[width=.4\textwidth]{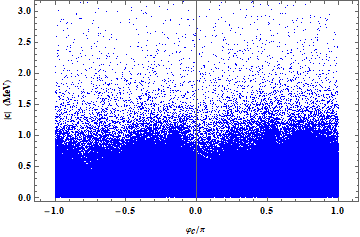} & \includegraphics[width=.4\textwidth]{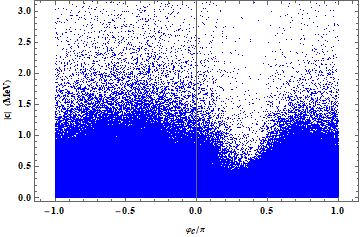} \\
    \includegraphics[width=.4\textwidth]{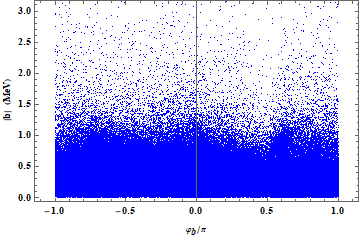} & \includegraphics[width=.4\textwidth]{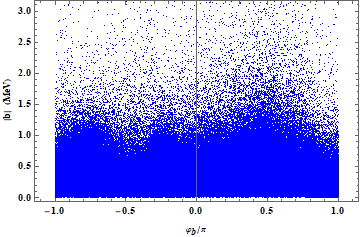}\\
\end{tabular}
\caption{Plot of the parameters of the model $|\kappa_1, |a|, |b|$ and $|c|$ versus their phases for NH (left panel) and IH (right panel).}
\end{figure}
%%%%%%%%%%%%%%%%%%%%%%%%%%%%%%%%%%%%5fig 4 %%%%%%%%%%
\begin{figure}[hbtp]
\centering
  \begin{tabular}{@{}cc@{}}
  \includegraphics[width=.4\textwidth]{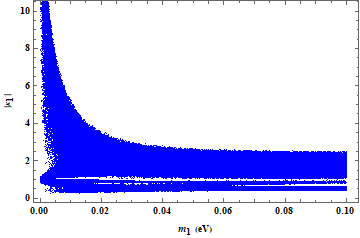} & \includegraphics[width=.4\textwidth]{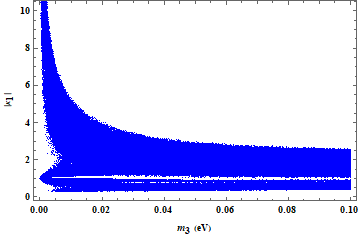} \\
    \includegraphics[width=.4\textwidth]{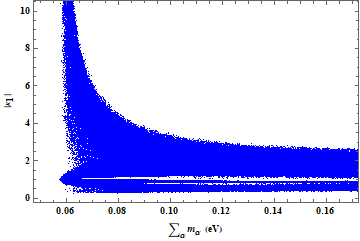} & \includegraphics[width=.4\textwidth]{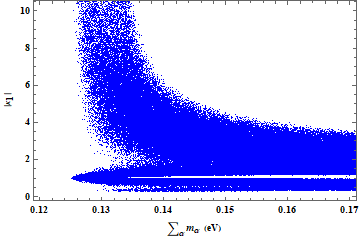} \\
%includegraphics[width=.4\textwidth]{fig4eNH.png} & \includegraphics[width=.4\textwidth]{fig4fIH.png} \\
\end{tabular}
\caption{Plot of the parameters of the model $|\kappa_1|$ versus the lightest neutrino mass, $\sum\limits_{\alpha=e,\mu,\tau} |m_{\nu_{\alpha}}|$  for NH (left panel) and IH (right panel).}
%\caption{Plot of the parameters of the model $|\kappa_1|$ versus the lightest neutrino mass, $\sum\limits_{\alpha=e,\mu,\tau} |m_{\nu_{\alpha}}|$ and $|m_{ee}|$ for NH (left panel) and IH (right panel).}
\end{figure}
%%%%%%%%%%%%%%%%%%%%%%%%%%%%%%%%%%%%%%%%%%%%%%%%%%%%%%%%%%%%%%%%%%%%%%%%%%%%%
%\newpage
%\mbox{}
%\newpage
\subsection{Baryogenesis via Resonant Leptogenesis} 
In our numerical analysis, we will consider scenarios with three nearly degenerate heavy right-handed Majorana neutrinos with masses at the TeV range. We will study the range of the model's parameters involved to reproduce the correct amount of the matter-antimatter asymmetry of the Universe. The observed baryon number asymmetry today is: 
\begin{eqnarray}
\eta_B & = & \frac{n_B - n_{\bar{B}}}{n_{\gamma}} = \left(6.04 \pm 0.08 \right) \times 10^{-10}
\end{eqnarray}
where $n_{B (\bar{B})}$ is the number of baryons (anti-baryons) density, $n_{\gamma}$ is the number density of photons. \\
As discussed earlier, we have expressed the neutrino Yukawa coupling $Y_{\nu}$ in terms of the model's parameters $\kappa_1, a, b, c, q$ and $r$. We fed the one million data set points $(\kappa_1, a, b, c, q,r)$, which are consistent with the low energy neutrino data, to compute the baryogenesis via resonant leptogenesis for each point of the data sets. In figures 5 and 6, we have plotted the baryon asymmetry $\eta_B$ against the model's parameters. Note that the observed baryon asymmetry is achieved (horizontal solid lines in figure 5) whithin the preferred regions of the model's parameters which is in the MeV or below. In figure 6, we show thw variation of the baryon asymmetry as a function of the phases $\varphi_{\kappa_1}, \varphi_a, \varphi_b$ and $\varphi_c$. The two top plots in figure 6 show that a successful leptoegenesis occurs when the phase of the scaling factor $\varphi_{\kappa_1}$ falls in the three disconnected regions. This is in good agreement with the global analysis at $3 \sigma$ level. It follows from the other plots in figure 6 that further constraints on the phases $\varphi_a, \varphi_b$  and $\varphi_c$  are required. \\

%%%%%%%%%%%%%%%%%%%%%%%%%%%%%%%%% fig 5 %%%%%%%%%%%%%%%%%%%%
\begin{figure}[hbtp]
\centering
  \begin{tabular}{@{}cc@{}}
  \includegraphics[width=.4\textwidth]{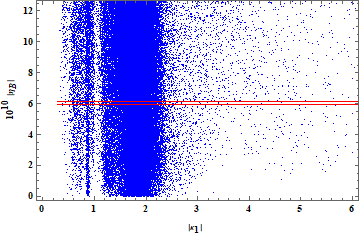} & \includegraphics[width=.4\textwidth]{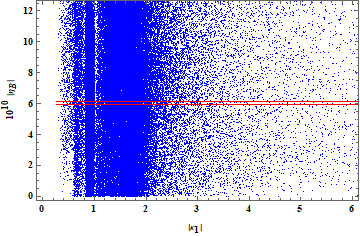} \\
    \includegraphics[width=.4\textwidth]{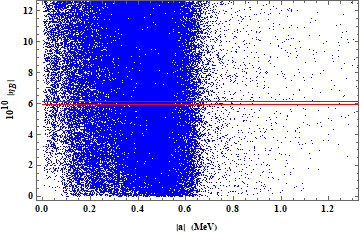} & \includegraphics[width=.4\textwidth]{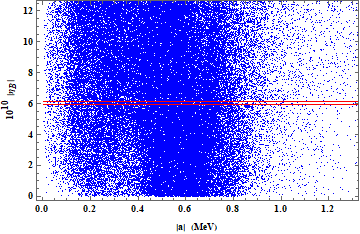} \\
    \includegraphics[width=.4\textwidth]{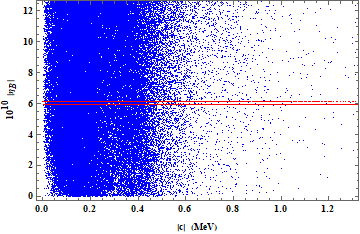} & \includegraphics[width=.4\textwidth]{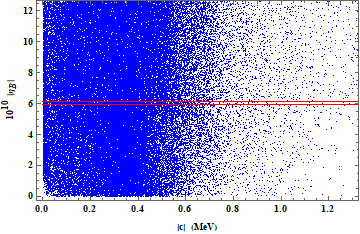} \\
    \includegraphics[width=.4\textwidth]{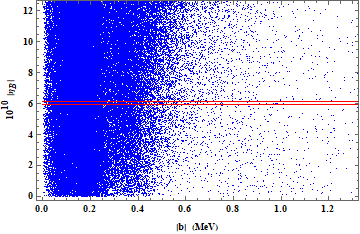} & \includegraphics[width=.4\textwidth]{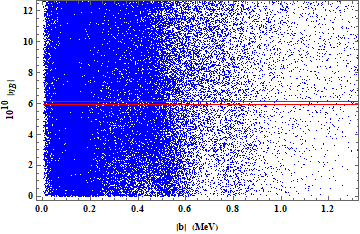} \\
\end{tabular}
\caption{Plot of the CP asymmetry $\eta_B$ versus the parameters of the model $|\kappa_1|, |a|, |c|, |d|$  and $r$ for NH (left panel) and IH (right panel).}
\end{figure}

%%%%%%%%%%%%%%%%%%%%%%%%%%%%% fig 6 %%%%%%%%%%%%%%%%%%%%%%%
\begin{figure}[hbtp]
\centering
  \begin{tabular}{@{}cc@{}}
  \includegraphics[width=.4\textwidth]{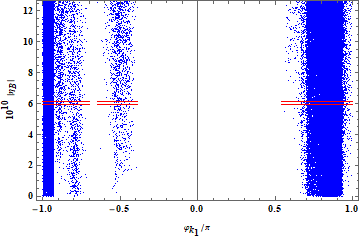} & \includegraphics[width=.4\textwidth]{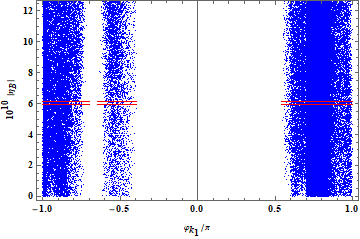} \\
    \includegraphics[width=.4\textwidth]{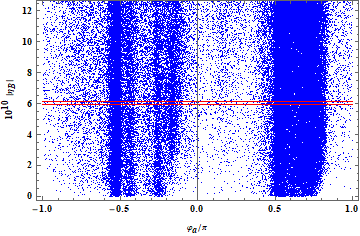} & \includegraphics[width=.4\textwidth]{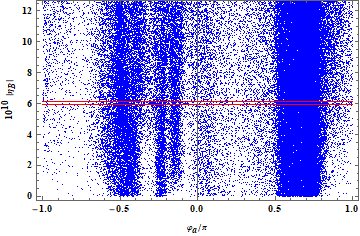} \\
    \includegraphics[width=.4\textwidth]{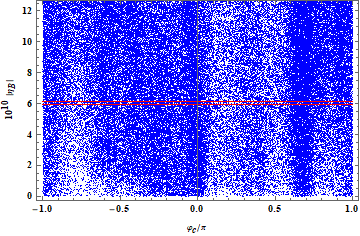} & \includegraphics[width=.4\textwidth]{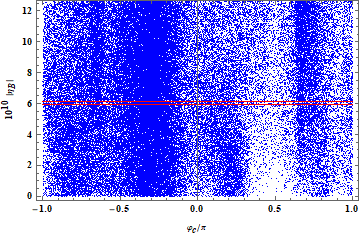} \\
    \includegraphics[width=.4\textwidth]{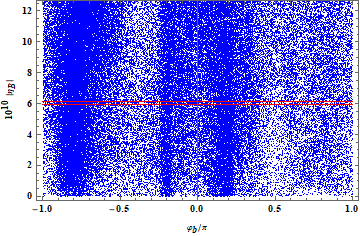} & \includegraphics[width=.4\textwidth]{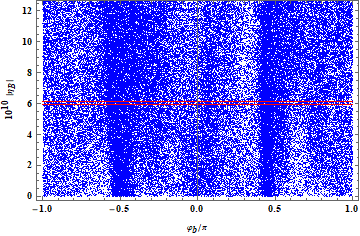} \\
\end{tabular}
\caption{Plot of the CP asymmetry $\eta_B$ versus the phases of the model for NH (left panel) and IH (right panel).}
\end{figure}

%%%%%%%%%%%%%%%%%%%%%%%%%%%%%%%%%%%%%%%%%%%%%%%%%%%%%%%%%%%%%
\subsection{Neutrinoless Double Beta Decay} 
Another implication of our model is the existence of
neutrinoless double-beta decay. The effective neutrino mass governing the $0 \nu 2 \beta$ decay is given by: 
\begin{eqnarray}
| m_{ee} | & = & | \sum_{i=1}^3 \mu_i U_{ei}^2| 
\end{eqnarray}
and corresponds to the (11) entry of the neutrino mass matrix $M_{\nu}$: 
\begin{eqnarray}
| m_{ee} | & = & | M_{\nu~11}|  \nonumber\\
& = & \frac{| (1-q r) \left(a^2 + 2 \kappa_1 \kappa_2 ~c ~d \right) - 
(r - q^2) \left( \kappa_2^2 c^2 + 2 \kappa_1 ~a ~b\right) +  
+ (q - r^2) \left( \kappa_1^2 b^2 + 2 \kappa_2 ~a ~c \right)|}{f (1- 3 q r +q^3 + r^3)}
\end{eqnarray}

The results for the baryon asymmetry $\eta_B$ as a function of $|m_{ee}|$ are depicted in figure 7. 

%%%%%%%%%%%%%%%%%%%%%%%%%%% fig 7 %%%%%%%%%%%%%%%
\begin{figure}[hbtp]
\centering
  \begin{tabular}{@{}cc@{}}
  \includegraphics[width=.4\textwidth]{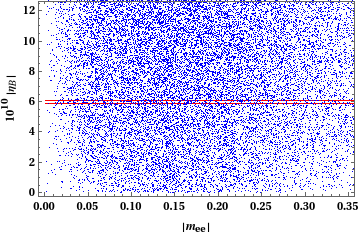} & \includegraphics[width=.4\textwidth]{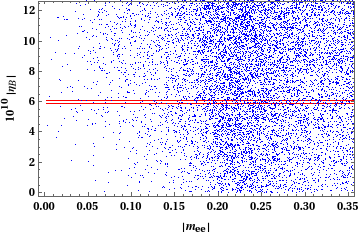} \\
\end{tabular}
\caption{Plot of the CP asymmetry $\eta_B$ versus $m_{ee}$ for NH (left panel) and IH (right panel).}
\end{figure}

The horizontal solid lines correspond to the allowed region of the observed baryon asymmetry. The important feature of the model is the existence of a lower bound $|m_{ee}| > 0.025~eV$ for NH and $|m_{ee}| > 0.10~eV$ for IH which is within the sensitivity reach of future double decay experiments~\cite{barabash}-\cite{licciardi}.
%%%%%%%%%%%%%%%%%%%%%%%%%%%%%%%%%%%%%%%%%%
\section{Conclusions}
We have proposed an $A_4$ flavor symmetry model to explain the baryon asymmetry of the universe. It is based on a type-I seesaw scenario with a retro-circulant heavy Majorana neutrino mass matrix and a scaling ansatz for the Dirac neutrino mass matrix. The heavy right-handed Majorana neutrinos are quasi-degenerate at TeV mass scale and the parameter space of the Dirac mass matrix elements lies near and below the MeV region. We have managed to reconstruct analytically all the elements of the Yukawa coupling matrix $Y_{\nu}$ from the low energy neutrino oscillation data. We have aslo shown that scaling ansatz on the neutrino Dirac mass matrix with the retro-circulant heavy Majorana mass matrix are consistent with the low energy neutrino data and the observed baryon asymmetry.
Finally, the model has an imperative implication on the allowed space for the effective Majorana neutrino mass $|m_{ee}|$ in order to account for the observed baryon asymmetry.
%=====================================

% The following MDPI journals use author-date citation: Arts, Econometrics, Economies, Genealogy, Humanities, IJFS, JRFM, Laws, Religions, Risks, Social Sciences. For those journals, please follow the formatting guidelines on http://www.mdpi.com/authors/references
% To cite two works by the same author: \citeauthor{ref-journal-1a} (\citeyear{ref-journal-1a}, \citeyear{ref-journal-1b}). This produces: Whittaker (1967, 1975)
% To cite two works by the same author with specific pages: \citeauthor{ref-journal-3a} (\citeyear{ref-journal-3a}, p. 328; \citeyear{ref-journal-3b}, p.475). This produces: Wong (1999, p. 328; 2000, p. 475)

%%%%%%%%%%%%%%%%%%%%%%%%%%%%%%%%%%%%%%%%%%
%% optional
%\sampleavailability{Samples of the compounds ...... are available from the authors.}

%% for journal Sci
%\reviewreports{\\
%Reviewer 1 comments and authors response\\
%Reviewer 2 comments and authors response\\
%Reviewer 3 comments and authors response
%}

%%%%%%%%%%%%%%%%%%%%%%%%%%%%%%%%%%%%%%%%%%
\end{document}